\documentclass[prd,eqsecnum,floatfix]{revtex4}

\usepackage{amsfonts}
\usepackage{amssymb}
\usepackage{graphicx}
\usepackage{pstricks}

\newcommand{\BOX}{\hbox {$\sqcap$ \kern -1em $\sqcup$}}

\newcommand{\be}{\begin{equation}}
\newcommand{\ee}{\end{equation}}
\newcommand{\ba}{\begin{eqnarray}}
\newcommand{\ea}{\end{eqnarray}}
\newcommand{\ban}{\begin{eqnarray*}}
\newcommand{\ean}{\end{eqnarray*}}

\newcommand{\bear}{\begin{eqnarray}}
\newcommand{\eear}{\end{eqnarray}}

\newcommand{\vev}[1]{\left\langle #1\right\rangle}

\newcommand{\lapproxeq}{\lower .7ex\hbox{$\;\stackrel{\textstyle
<}{\sim}\;$}}
\newcommand{\gapproxeq}{\lower .7ex\hbox{$\;\stackrel{\textstyle
>}{\sim}\;$}}
\newcommand{\stackdown}[2]{\lower 1.4ex\hbox{$\;\stackrel{\textstyle{#1}}
{\scriptstyle{#2}}\;$}}
\newcommand{\beq}{\begin{equation}}
\newcommand{\eeq}{\end{equation}}
\newcommand{\lsp}{\tilde{\chi}}

\begin{document}

\rightline{ACT-06-06, MIFP-06-20}

\title{Dilaton and off-shell (non-critical string) effects
in Boltzmann equation for species abundances}

\author{A.B. Lahanas}
\affiliation{University of Athens, Physics Department,
Nuclear and Particle Physics Section,  GR157 71,
Athens, Greece.}

\author{N.E. Mavromatos}
\affiliation{King's College London, University of London,
Department of Physics, Strand WC2R 2LS, London, U.K.}

\author{D.V. Nanopoulos}
\affiliation{George P.\ and Cynthia W.\ Mitchell Institute for
Fundamental
Physics, Texas A\&M
University, College Station, TX 77843, USA; \\
Astroparticle Physics Group, Houston
Advanced Research Center (HARC),
Mitchell Campus,
Woodlands, TX~77381, USA; \\
Academy of Athens,
Division of Natural Sciences, 28~Panepistimiou Avenue,
Athens 10679,
Greece}

\begin{abstract}

In this work we derive the modifications to the Boltzmann equation
governing the cosmic evolution of relic abundances induced by
dilaton dissipative-source and non-critical-string terms in
dilaton-driven non-equilibrium string Cosmologies. We also discuss
briefly the most important phenomenological consequences, including
modifications of the constraints on the available parameter space of
cosmologically appealing particle physics models, imposed by recent
precision data of astrophysical measurements.

\end{abstract}


\maketitle

\section{Introduction}

In a previous work~\cite{dglmn} we have discussed a case study of
dissipative Liouville-string cosmology, involving non critical
string cosmological backgrounds, with the identification~\cite{emn}
of target time with the world-sheet zero mode of the Liouville
field~\cite{ddk}. Such cosmologies were found to asymptote (in
cosmic time) with the conformal backgrounds of \cite{aben}, which
are thus viewed as equilibrium (relaxation) configurations of the
non-equilibrium cosmologies.

In terms of microscopic considerations, a model for such departure
from equilibrium could be considered the collision of two brane
worlds, which results~\cite{brany} in departure from conformal
invariance of the effective string theory on both the bulk and the
brane world, and thus in need for Liouville dressing to restore this
symmetry~\cite{ddk}. Dynamical arguments~\cite{gravanis}, then,
stemming from minimization of effective potentials in the low-energy
string-inspired effective field theory on the brane, imply the
(eventual) identification of the zero mode of the Liouville mode
with (a function) of target time~\cite{emn}. There is an inherent
\emph{time irreversibility} in the process, which is associated with
basic properties of the Liouville mode, viewed as a local
(dynamical) renormalization group (RG) scale on the world-sheet of
the string. This implies \emph{relaxation} of the associated dark
energy of such cosmologies, and a gravitational friction, associated
with the conformal theory central charge deficit. For a recent
review we refer the reader to \cite{brany}, where concepts and
methods are outlined in some detail.

For our purposes in this letter we would like to concentrate on one
interesting aspect of the non-critical string cosmologies,
associated with the off-equilibrium effects on the Boltzmann
equation describing relic abundances and the associated
particle-physics phenomenology. Indeed, in conventional cosmologies,
a study of relic abundances by means of Boltzmann equation that
governs their cosmic time evolution yields important
phenomenological constraints on the parameters of particle physics
(supersymmetric) models using recent (WMAP~\cite{wmap} and other)
astrophysical Cosmic Microwave Background (CMB) data.
Essentially, the astrophysical data constraint severely in
some cases the available phase-space distributions of favorite
supersymmetric dark matter candidates such as
neutralinos~\cite{susyconstr}. When off-equilibrium, non-critical
string cosmologies are considered~\cite{dglmn}, which notably are
consistent with the current astrophysical data from supernovae, as
demonstrated recently~\cite{mitsou2}, then there are significant
modifications to the Boltzmann equation, stemming from extra sources
(dilaton) and off-shell (non-critical, non equilibrium) terms, which
affect the time evolution of the phase-space density of the species
under consideration.

It is the purpose of this paper to derive such modifications in
detail, and then use them in order to discuss briefly
particle-physics models constraints, especially from the point of
view of supersymmetry. With regards to this last issue, in this
article we shall present an approximate analytical treatment, which
will only provide hints to what may actually happen. A complete
analysis requires numerical studies which are postponed for a future
publication.

\section{Modified Boltzmann Equation in Non-critical Strings}

Consider the phase space density of a species X, which is assumed coupled to the off-shell (non-critical string) background terms:
$$ f(|\vec{p}|, t) $$
where quantities refer to the Einstein frame, and the Einstein
metric is assumed to be of Robertson-Walker (RW) type. Throughout
this work we follow the normalization and conventions of
\cite{aben}.

For completeness we state the main relationships between the
Einstein and $\sigma$-model frames~\cite{aben}: \ba g_{\mu\nu} =
e^{-2\Phi} g_{\mu\nu}^\sigma ~, \qquad \frac{\partial t}{\partial
t^\sigma} = e^{-\Phi} \label{frames} \ea where $\Phi $ is the
dilaton field, and the superscript $\sigma$ denotes quantities
evaluated in the $\sigma$-model frame.

Since in the RW Einstein frame $g_{00}=-1$, and in the cosmic Einstein co-moving frame we have for a generic massive species with mass $m$ (that we shall be interested in in this work)
$p_\sigma ^\mu = m dx^\mu/d\tau = m (dx^\mu/dt )(dt/dt_\sigma )= mp^\mu e^{-\Phi}$, it can be readily seen that
\ba
|\vec{p}^{~\sigma} | \equiv
\left(p^{i, \sigma}p^{j, \sigma} g_{ij}^\sigma\right)^{1/2} =
\left(p^i p^j g_{ij}\right)^{1/2} \equiv
|\vec{p} | = a(t) \left(\sum_{i=1}^{3} p^i p^i \right)^{1/2}
\label{p2frames}
\ea

We shall be interested in the action of the relativistic Liouville
operator ${\widehat L}$, acting on a phase-space density of species
X, $f(|\vec{p}|, t, g^i)$, which in general depends on the off-shell
backgrounds $g^i = (g_{ii}, \Phi )$, where $g_{ii}$ are the spatial
components of the RW space-time metric in Einstein frame, and $\Phi$
is the dilaton. The off-shell dependence comes about due to the
interpretation of target time as a local world-sheet
renormalization-group scale, the Liouville field.

We commence our analysis by recalling the relativistic form of the
Liouville operator in conventional general relativity: \ba {\widehat
L}[f] = \left(p^\alpha \frac{\partial}{\partial x^\alpha} + \Gamma
^\alpha _{\beta\sigma} p^\beta p^\sigma \frac{\partial}{\partial
p^\alpha}\right)f \label{liouvop} \ea The second term on the r.h.s.
is the relativistic form of the force, following from the geodesic
equation. For a RW Universe, only the time-energy part survives from
the first term, i.e. \ba p^\alpha \frac{\partial}{\partial x^\alpha}
= E \frac{\partial}{\partial t}~. \label{liouvop2} \ea Moreover, the
connection part receives non trivial contributions only from the
terms $\sum_{i}\Gamma_{ii}^0 p^i p^i \frac{\partial}{\partial E}$,
$i=1,2,3$ a spatial index (assuming for concreteness an already
compactified string theory, or a theory on a three-brane world),
with $\Gamma_{ii}^0 = -a {\dot a}$, where the overdot denotes
derivative w.r.t.\ cosmic RW time $t$, identified in our string
theory with the Einstein-frame RW time (\ref{frames}).

Hence, the conventional part of the Liouville operator in a RW Universe
would read~\cite{kolb}:
\begin{equation}
{\widehat L}_{\rm conv}
= E\frac{\partial}{\partial t} - a{\dot a}\sum_{i=1}^{3}
p^i p^i \frac{\partial}{\partial E} = E\frac{\partial}{\partial t}
- \frac{{\dot a}}{a}\sum_{i=1}^{3}
|\vec p |^2\frac{\partial}{\partial E}
\label{convliouvop}
\end{equation}

The presence of off-shell, non-critical (Liouville)
string~\cite{ddk} backgrounds, upon the (dynamical)
identification~\cite{emn,gravanis,brany} of the Liouville mode with
the target time, one will receive extra contributions in the
expression for the associated Liouville operator, stemming from the
fact that the latter is nothing but a total time derivative.

In the context of our string discussion it is important to specify
that the initial frame, from which we commence our discussion, is
the so-called Einstein frame. It is in this frame that the usual RW
cosmology is obtained in string theory~\cite{aben}, for which the
expression (\ref{convliouvop}) is valid.

To discuss the non-critical-string (off-shell) corrections to
Boltzmann equation, it will be necessary to consider time
derivatives in the $ \sigma$-model frame. This is due to the fact
that it is in this frame that the target time $X^0 \equiv t_\sigma $
is related simply to the Liouville mode $\varphi $ in non-critical
string theories~\cite{emn},
\begin{equation}
\varphi + t_\sigma = 0~. \label{liouv-time}
\end{equation}
This relation is obtained dynamically from minimization arguments of
the low-energy effective potential of some physically interesting
cosmological models, for instance those involving colliding brane
worlds~\cite{gravanis,brany}. To be precise, the initial relation
(\ref{liouv-time}) derived in the specific model of \cite{gravanis}
reads: $\varphi/\sqrt{2} + t_\sigma = 0~$, with the factor of
$\sqrt{2}$ arising from (logarithmic) conformal-field-theory
considerations, and being crucial for yielding a Minkowski
space-time in the scenario of \cite{gravanis}, where the coordinate
$X^0$ assumed an initial Euclidean signature, appropriate for a
world-sheet path-integral quantization. Nevertheless, in our
approach below we shall absorb the $\sqrt{2}$ into the normalization
of the units of the Regge slope $\alpha '$, for convenience.

In such scenarios the target-space dimensionality of the string is
extended to $D+1$ initially, with two time-like coordinates, $t$ and
$\varphi$. Eventually, these two coordinates are identified (c.f.
(\ref{liouv-time})) in the solution of the generalized conformal
invariance conditions that express the restoration of the conformal
invariance by the Liouville mode, as we discuss below. This implies
that, initially, the Liouville operator (\ref{liouvop}) acquire
extra Liouville components $(\varphi, p^\varphi) $, with $p^\varphi
\equiv E_\varphi $. Hence, there are additional structures on the
right hand side of (\ref{liouvop}), of the form $E_\varphi
\frac{\partial }{\partial \varphi }$, together with the
corresponding amendments in the Christoffel-symbol dependent
$\partial/\partial p^\varphi$ terms, stemming from the extra
Liouville coordinate on the target space time, which implies
appropriate extra components of the extended-target-space-time
metric. The (dynamical) implementation of (\ref{liouv-time})
restricts oneself to a $D$-dimensional hypersurface in the extended
target-space, with the identification of $E_\varphi \to E$, the
ordinary (physical) energy. From now on we restrict ourselves on
this hypersurface, bearing in mind however that the identification
of the Liouville mode $\varphi$ with the target time should only be
implemented at the end, and thus any dependence on $\varphi$ should
be kept explicit at intermediate steps of the pertinent
calculations.

The connection between time derivatives in the Einstein and
stringy-$\sigma$-model frames is provided by the chain rule of
differentiation $\frac{\partial }{\partial t_\sigma} =
\frac{\partial t}{\partial t_\sigma}\frac{\partial }{\partial t}$,
using (\ref{frames}). This will result in multiplicative factors of
$e^{-\Phi}$ in front of the appropriate non-critical-string
modifications of the Liouville operator (\ref{convliouvop}).

Since, as mentioned above, the Liouville operator is essentially a
total  time derivative operator, and in our case time is related to
a world-sheet renormalization group scale, $\rho (\sigma, \tau )$,
taken to be local on the two-dimensional surface as a result of
world-sheet general covariance~\cite{osborn}, the sought-for
non-critical-string modifications of the Boltzmann equation emerge
from the implicit dependence of the $g^i$ background fields on $\rho
(\sigma,\tau)$, that is the corresponding $\beta$-functions, which
in a critical-string theory  would vanish.

We now remark that in the approach of \cite{emn}, the local RG scale
$\rho(\xi)$ was identified with the dynamical Liouville mode
$\varphi$, which implied that in this approach the local conformal
invariance of the world-sheet of the string was restored~\cite{ddk},
\begin{equation}
\rho \equiv \varphi ~.
\label{rhovarphiident}
\end{equation}
With these in mind we then modify the relativistic form
(\ref{liouvop}), (\ref{liouvop2}) by replacing \ba &&
\frac{\partial}{\partial t} \to \frac{d}{d t} \equiv \frac{\partial
}{\partial t} + \int_\Sigma \frac{\partial t_\sigma}{\partial
t}\frac{\partial \rho (\xi)}{\partial t_\sigma} \frac{\partial
g^{``i''}}{\partial \rho (\xi)} \frac{\partial }{\partial
g^{`` i ''} }   = \nonumber \\
&& \frac{\partial }{\partial t} + \eta
e^\Phi~\beta^{``i''}\frac{\partial }{\partial g^{`` i ''} }~,
\label{totalliouvope} \ea where $\xi = \sigma, \tau$, denotes the
world-sheet coordinates, $\int _\Sigma $ is a world-sheet
integration, and the index $``i''$ runs on both, a (discrete)
background field space, $\{ g_{ii}(y), \Phi (y)\}, i=1,2,3 $ and a
continuous $D$-dimensional (target) space-time $y$. Hence, summation
over $``i''$ includes integration $\int d^Dy\sqrt{-g} (...)~.$ Note
that since the (physical) target time in the $\sigma$-model frame,
$t_\sigma$, is only the world-sheet zero mode of the corresponding
$\sigma$-model field $X^0(\xi)$, eventually the world-sheet
integration $\int_\Sigma$ in (\ref{totalliouvope}) disappears, since
only the world-sheet zero mode $\rho$ of the local RG scale $\rho
(\xi)$ (Liouville mode) will yield a non zero contribution. In the
last equality of the right-hand side of (\ref{totalliouvope}), the
quantity $\eta $ is defined as: $\eta \equiv \frac{\partial
\rho}{\partial t_\sigma}$. If we use the identification $\rho (\xi)
= \varphi (\xi ) $ and Eq.~(\ref{liouv-time}), then $\eta = -1$, but
at this stage we keep it general to demonstrate explicitly the
renormalization-scheme dependence (i.e. choice of the local scale
$\rho (\xi )$).

The non-critical string contributions $\frac{\partial g^i}{\partial
\rho } = {\tilde \beta}^i$ are the Weyl anomaly coefficients of the
string, but with the renormalized $\sigma$-model couplings $g^i$
being replaced by the Liouville-dressed~\cite{ddk} quantities. This
is a particular feature of the approach of \cite{emn} viewing the
Liouville mode as a local world-sheet RG scale, as mentioned
previously. In turn the latter is also identified (c.f.
(\ref{liouv-time})) with (an appropriate  function of) the target
time $t$.

The dynamics of this latter identification is encoded in the
solution of the generalized conformal invariance conditions, after
Liouville dressing, which read in the $\sigma$-model
frame~\cite{emn,ddk}:
\begin{equation}
-{\tilde \beta^i } = {g^i}'' + Q{g^i}' \label{liouveq}
\end{equation}
where the prime denotes differentiation with respect to the
Liouville zero mode $\rho$, and the overall minus sign on the
left-hand side of the above equation pertains to supercritical
strings~\cite{aben}, with a time like signature of the Liouville
mode, for which the central charge deficit $Q^2 > 0$ by convention.

Notice the dissipation, proportional to the (square root) of the
central charge deficit $Q$, on the right-hand side of
(\ref{liouveq}), which heralds the adjective \emph{Dissipative} to
the associate non-critical-string-inspired Cosmological model.
Moreover, the Weyl anomaly coefficients ${\tilde \beta^i },~i=
\{\Phi, g_{\mu\nu} \}$, whose vanishing would guarantee local
conformal invariance of the string-cosmology background, are
associated with off-shell variations of a low-energy effective
string-inspired action, ${\cal S}[g^j]$
\begin{equation}
\frac{\delta {\cal S}[g] }{\delta g^i} = {\cal G}_{ij}{\tilde
\beta}^j \qquad {\cal G}_{ij}= {\rm Lim}_{z \to 0} z^2 {\overline z}
\langle V_i(z,{\overline z}) V_j(0,0)\rangle
 \label{offshell}
\end{equation}
with $z,{\overline z}$ (complex) world-sheet coordinates, ${\cal
G}_{ij} $ the Zamolodchikov metric in theory space of strings, and
$V_i$ the $\sigma$-model vertex operators associated with the
$\sigma$-model background field $g^i$. It is this off-shell relation
that characterizes the entire non-critical ($Q$) Cosmology
framework, associated physically with a non-equilibrium situation as
a result of an initial cosmically catastrophic event, at the
beginning of the (irreversible) Liouville/cosmic-time
flow~\cite{brany}.

The detailed dynamics of (\ref{liouveq}) are encoded in the solution
for the scale factor $a(t)$ and the dilaton $\Phi$ in the simplified
model considered in \cite{dglmn}, after the identification of the
Liouville mode with the target time (\ref{liouv-time}). In fact,
upon the inclusion of matter backgrounds, including dark matter
species, the associated equations, after compactification to four
target-space dimensions,  read in the Einstein frame~\cite{dglmn}:
\ba &&3 \; H^2 - {\tilde{\varrho}}_m - \varrho_{\Phi}\;=\;
\frac{e^{2 \Phi}}{2} \; \tilde{\cal{G}}_{\Phi} \nonumber  \\
&&2\;\dot{H}+{\tilde{\varrho}}_m + \varrho_{\Phi}+
{\tilde{p}}_m +p_{\Phi}\;=\; \frac{\tilde{\cal{G}}_{ii}}{a^2} \nonumber  \\
&& \ddot{\Phi}+3 H \dot{\Phi}+ \frac{1}{4} \; \frac{\partial {
\hat{V}}_{all} }{\partial \Phi} + \frac{1}{2} \;(
{\tilde{\varrho}}_m - 3 {\tilde{p}}_m )= - \frac{3}{2}\; \frac{
\;\tilde{\cal{G}}_{ii}}{ \;a^2}- \, \frac{e^{2 \Phi}}{2} \;
\tilde{\cal{G}}_{\Phi}  \; .\label{eqall} \ea
where ${\tilde{\varrho}}_m$ (${\tilde{p}}_m $) denotes the matter
energy density (pressure), including dark matter contributions, and
$\varrho_{\Phi}$ ($p_\Phi$) the corresponding quantities for the
dilaton dark-energy fluid.
All derivatives in (\ref{eqall}) are with respect the Einstein time $t$ which is related to
the Robertson-Walker cosmic time $t_{RW}$ by $t=\omega \;t_{RW}$.  Without loss of generality
we have taken $\omega = \sqrt{3} H_0$ where $H_0$ is the Hubble constant. With this choice
for $\omega$ the densities appearing in (\ref{eqall}) are given in units of the critical density. Then
one can see for instance that if the time $t_{RW}$ is used the first of equations (\ref{eqall}) above receives its
familiar form in the RW geometry when the contributions of the dilaton and non-critical terms are absent.

The overdots in the above equations denote derivatives with respect
to the Einstein time. Their right-hand side contain the non-critical
string {\it off-shell} terms: \ba && \tilde{\cal{G}}_{\Phi} \;=\;
e^{\;-2 \Phi}\;( \ddot{\Phi} - {\dot{\Phi}}^2 + Q e^{\Phi}
\dot{\Phi}) \nonumber \\ && \tilde{\cal{G}}_{ii} \;=\; 2 \;a^2 \;(\;
\ddot{\Phi} + 3 H \dot{\Phi} + {\dot{\Phi}}^2 + ( 1 - q ) H^2 + Q
e^{\Phi} ( \dot{\Phi}+ H )\;) \; .  \ea
In the above equations $H=~(\dot a)/a$ is the Hubble parameter and
$q$ is the deceleration parameter of the Universe $q \equiv -
\ddot{a} a / {\dot{a}}^2$, and are both functions of (Einstein
frame) cosmic time. The potential $\; {\hat{V}}_{all} $ appearing
above is defined by $\; {\hat{V}}_{all}= 2 Q^{\;2} \exp{\;( 2 \Phi
)}+ V \;$ where, in order to cover more general cases, we have also
allowed for a potential term in the four-dimensional action  $- \int
d^4y \sqrt{-G}\;V$ in addition to that dependent on the central
charge deficit term $Q$. Although we have assumed a (spatially) flat
Universe, the terms on the r.h.s., which manifest departure from the
criticality, act in a sense like curvature terms as being non-zero
at certain epochs. The dilaton energy density and pressure are given
in this class of models by:
\ba && \varrho_{\Phi}\;=\;\frac{1}{2}\;( \; 2
{\dot{\Phi}}^2+{\hat{V}}_{all} ) \nonumber \\
&&p_{\Phi}\;=\;\frac{1}{2}\;( \; 2 {\dot{\Phi}}^2-{\hat{V}}_{all} )
\; \; .  \ea
Notice that the dilaton field is not canonically normalized in this
convention and its dimension has been set to zero.

For completeness, we mention at this point that the dependence of
the central charge deficit $Q(t)$ on the cosmic time stems from the
running of the latter with the world-sheet RG
scale~\cite{emn,brany}, and is provided by the Curci - Paffuti
equation~\cite{curci} expressing the renormalizability of the
world-sheet theory. To leading order in an $\alpha '$ expansion,
which we restrict ourselves in \cite{dglmn} and here, this equation
in the Einstein frame reads:
\ba \frac{d \tilde{\cal{G}}_{\Phi} }{d t_E} \;=\; - 6\; e^{\;-2
\Phi}\;( H + \dot{\Phi} ) \; \frac{ \;\tilde{\cal{G}}_{ii}}{ \;a^2}
\; .  \label{CXP} \ea
For future use we also state here the corresponding continuity of
the matter stress tensor, which is not an independent equation, but
can be obtained from (\ref{eqall}) by appropriate algebraic
manipulations:
\ba \frac{d {\tilde{\varrho}}_m }{dt_E}+ 3 H (
{\tilde{\varrho}}_m +{\tilde p}_m) + \frac{\dot Q}{2} \frac{\partial
{ \hat{V}}_{all} }{\partial Q} - \dot{\Phi}\;({\tilde{\varrho}}_m -
3 {\tilde{p}}_m )\;=\; 6\;(H+\dot{\Phi})\;
\frac{\;\tilde{\cal{G}}_{ii}}{a^2} \; . \label{contin} \ea This
expresses the non-conservation equation of matter as a result of its
coupling to \emph{both}, the dilaton source terms and the off-shell,
non-equilibrium (non-critical-string) backgrounds.

A consistent solution of $a(t)$, $\Phi(t)$, and the various
densities, including back reaction of matter onto the space-time
geometry, has been discussed in \cite{dglmn}, where we refer the
interested reader for further study. Also note that a preliminary
comparison of such non-critical strings theories with astrophysical
data, demonstrating consistency at present, is given in
\cite{mitsou2}.

After this necessary digression we now come back to discussing the
derivation of the dilaton-source and non-critical-string induced
modifications to the Boltzmann equation, governing the cosmic
evolution of the various species densities. It is important for the
reader to bear in mind already at this stage that the Boltzmann
equation does not contain any independent information from the
dynamical equations (\ref{eqall}), but it should be rather viewed as
an effective way of describing the cosmic evolution of the density
of a given species, consistent with the continuity equation
(\ref{contin}) for the total matter energy density. We shall come
back to this important point later on.

At the moment, let us concentrate first on the dilaton-source and
non-critical-string background contributions to the Liouville
operator (\ref{totalliouvope}). The presence of (time-dependent)
dilaton source terms implies an explicit dependence of the phase
space density of a species $f$ on $\Phi$, while the non-conformal on
the world-sheet) nature of the metric and dilaton background, induce
a Liouville mode $\rho$ dependence through the corresponding
backgrounds:
\begin{equation}
f(|\vec p|, t, \Phi (t,\rho), g_{\mu\nu}(t,\rho))~.
\label{fextra}
\end{equation}
The non-critical string terms can be expressed, as we have seen
above, in terms of the corresponding Weyl anomaly coefficients,
which are non zero as a result of departure from conformal
invariance of the pertinent string background. Despite their
non-linear looking appearance when expressed in terms of the Weyl
anomaly coefficients (which depend on the Ricci tensor and second
covariant derivatives of the Dilaton field), such terms acquire a
particularly simple linear form once the identification of time with
the Liouville mode (\ref{liouv-time}) is implemented, which in
effect implies that the explicit solution of (\ref{liouveq}),
(\ref{eqall}) must be taken into account when discussing the
Boltzmann equation.

With the above in mind, the form of the non-critical-string
(off-shell) and dilaton-source corrections to the Liouville operator
reads (expressing quantities in the Einstein frame, taking into
account the fact that in the $\sigma$-model frame
$g_{\mu\nu}^\sigma$ and $\Phi$ are treated as independent
field-theory variables, and using (\ref{frames})): \ba && {\widehat
L}_{\rm off-shell,~\rm dil} \equiv E \dot \Phi \frac{\partial
}{\partial \Phi} + E \eta e^\Phi {\tilde \beta}^{``i''}
\frac{\partial}{\partial g^{``i''}} = E \dot \Phi \frac{\partial
}{\partial \Phi}+ E \eta e^\Phi \left(\sum_{i=1}^3 \frac{\partial
g_{ii}^\sigma}{\partial \rho} \frac{\partial g_{ii}}{\partial
g_{ii}^\sigma } \frac{\partial }{\partial g_{ii}} + \frac{\partial
\Phi}{\partial \rho} \frac{\partial }{\partial \Phi}\right) =
\nonumber \\
&& E \dot \Phi \frac{\partial }{\partial \Phi} + \eta E e^\Phi \{ 2
a^2 \left(\frac{a'}{a} + {\Phi '}\right)\sum_{i=1}^3 \frac{\partial
}{\partial g_{ii}}  +  {\Phi '} \frac{\partial }{\partial \Phi}
\}\ea where the overdot denotes the derivative with respect to the
(Einstein frame) cosmic time $t$, and the prime the derivative with
respect to $\rho$, and we used the fact that:
\ba
\frac{\partial g_{ii}^\sigma}{\partial \rho} = 2 \; a^2\;
e^{2 \Phi}\;( \frac{a '}{a}+\Phi ') \; \;,\; \;
\frac{\partial g_{ii}}{\partial g_{ii}^\sigma} = e^{- 2 \Phi}
\ea
with $g_{00} = -1, ~g_{ii} =
a^2$, and $\frac{\partial g_{ii}^\sigma }{\partial \rho} = {\tilde
\beta_{ii}^{\rm Grav}}$ the corresponding Weyl anomaly coefficient,
in order to express ${\tilde \beta_{ii}^{\rm Grav}}$ in terms of
quantities in the Einstein frame. We have also been careful to
separate the dilaton
dependence on $\rho (t)$ from that on the ordinary time $t$ (prior to the
identification (\ref{liouv-time})). We remind
the reader, at this point, that
the identification (\ref{liouv-time}) should only be applied
at the end of the calculations, and should be viewed as
a specific choice of ``renormalization''
scheme. The ``renormalization'' scheme (\ref{liouv-time}),
i.e. $\eta = -1$, which is obtained by a physical principle
(dynamics) in certain models of interest to us here~\cite{gravanis},
may thus be viewed in this context as the ``physical scheme'', where
observational cosmology can be analyzed.
In such a scheme there is a simplification of the
complicated non-linear expressions encoded in the Weyl anomaly
coefficients.

Taking into account the implicit dependence of $f(|\vec p|, t,
\Phi)$ on $g_{ii}$ through $|{\vec p}| = \left( \sum_{i=1}^3 p^i p^i
g_{ii}\right)^{1/2}$, it is evident that $\partial f/\partial g_{ii}
= \left(\partial |\vec p|/\partial g_{ii} \right)\partial f / \partial
|\vec p| = \frac{p_i p_i }{2|\vec p|}\frac{\partial f}{\partial |\vec
p|}$, from which we can easily see that the final form of the action
of the Liouville operator, including dilaton/source terms and
non-critical string corrections, on the phase-space density $f$
(\ref{fextra}), is
such that: \ba && \left({\widehat L}_{\rm conv} + {\widehat L}_{\rm
off-shell,dil}\right)f =
C[f]~, \nonumber \\
&& \frac{\partial f}{\partial t} = \frac{\dot a}{a} \frac{|\vec
p|^2}{E}\frac{\partial f}{\partial E} -\eta e^\Phi
\left(\frac{a'}{a} + \Phi ' \right)|\vec p|\frac{\partial
f}{\partial |\vec p|} - \left(\dot \Phi + \eta e^\Phi {\Phi
'}\right)\frac{\partial f}{\partial \Phi} + \frac{1}{E}C[f]~. \ea
Upon considering the action of the above operator on the density of
a given species $X$, $n \equiv \int d^3 p f $ we then arrive, after
some straightforward momentum integration by parts, at the
\emph{modified} Boltzmann equation for a four-dimensional effective
field theory after string compactification (or restriction on
three-brane worlds), in the presence of non-critical (off-shell)
string backgrounds and dilaton source terms: \ba && \frac{d n}{d t}
= \frac{\dot a}{a} \int d^3p \frac{|\vec p|^2}{E}\frac{\partial
f}{\partial E} -\left(\dot \Phi + \eta e^\Phi {\Phi '}\right) \int
d^3p \frac{\partial f}{\partial \Phi} -
\eta e^\Phi \left(\frac{a'}{a} + {\Phi '}\right)\int d^3p |\vec p|\frac{\partial f}{\partial |\vec p|} + \int d^3p \frac{C[f]}{E}~ \Longrightarrow \nonumber \\
&& \frac{d n}{dt} + 3\frac{\dot a}{a}n = 3\eta e^\Phi
\left(\frac{a'}{a} + \Phi' \right)n + \int \frac{d^3p}{E}C[f] -
\left(\dot \Phi + \eta e^\Phi {\Phi '}\right) \int d^3p
\frac{\partial f}{\partial \Phi} \label{liouvtotal} \ea The
collision term $C[f]$ assumes the usual form in conventional
particle cosmology~\cite{kolb}. The reader is invited to compare the
final equation for the cosmic time evolution of the density $n$ in
the second line of (\ref{liouvtotal}) to the continuity equation for
the total energy density of matter (\ref{contin}). As discussed
previously, the Boltzmann equation (\ref{liouvtotal}) should be
compatible in the sense of leading to no extra information) with the
conservation equation (\ref{contin}), as well as the (modified)
Einstein equations (\ref{eqall}).

In the above scenario, where the non-critical string contributions
have been obtained through the identification of time with the
(world-sheet zero-mode of the) Liouville field $\rho$
(\ref{liouv-time}), there is some \emph{universality} in the
coupling of \emph{all} matter species, including dark matter, to
these off-shell background terms, which may be traced to the
equivalence of all species coupled to gravity. This is to be
contrasted with the more general phenomenological case studied in
\cite{dglmn}, where scenarios involving only the  coupling of the
non-critical string backgrounds with the exotic dark matter species
have been considered as well.

The final issue to be discussed in this section pertains to the form
of the dependence of $f$ on the dilaton source terms, which would
survive a dilaton-driven critical-string cosmology case, such as the
one considered in \cite{veneziano}. We constrain this form by
requiring that in the Einstein frame there are two types of dependence
on $\Phi$: (i) \emph{explicit}, of
the form $e^{-4\Phi}$, arising from the fact that
in our approach, the phase space density is constructed
as a quantity in the $\sigma$-model frame of the string, which is then
expressed in terms of quantities in the Einstein frame.
As such, it is by definition (as a density) \emph{inversely}
proportional to the proper
$\sigma$-model frame volume $V^\sigma = \int d^4x \sqrt{-g^\sigma}
\propto e^{4\Phi}$ on account of (\ref{frames});
(ii) \emph{implicit}, corresponding to a
dependence on $\Phi$ through the Einstein-frame metric $g_{ii}$
(\ref{frames}). Hence the general structure of $f$ is
of the form:
\begin{equation} \label{structuref}
f (\Phi , \vec p, \vec x, g_{\mu\nu}^\sigma =e^{2\Phi}g_{\mu\nu};~t ) \propto
e^{-4\Phi}{\cal F}(|\vec p|, \vec x, t)
\end{equation}
This implies that: \ba && \int d^3p \frac{\partial
f}{\partial \Phi} = -4\int d^3p f +  \sum_{i=1}^3 \int d^3p \frac{\partial
g_{ii}}{\partial \Phi}\frac{\partial f}{\partial g_{ii}} = -4n -2\int
d^3p \sum_{i=1}^3 g_{ii}\frac{\partial |\vec p|}{\partial
g_{ii}}\frac{\partial f}{\partial |\vec p|} = \nonumber \\
&& -4n -\int d^3p |\vec p|\frac{\partial f}{\partial |\vec p|}= -4n + 3\int
d^3p f(|\vec p|,t) = -n~, \label{dilatondepen} \ea
where in the last step we have performed appropriate partial (momentum-space)
integrations.

The final form of the Liouville operation (\ref{liouvtotal}), then,
reads:
\ba \frac{d n}{dt} + 3\left(\frac{\dot a}{a}\right)n - \dot \Phi n
= 3\eta e^\Phi\frac{a'}{a}n + 4\eta e^\Phi \Phi ' n
+ \int \frac{d^3p}{E}C[f]
\label{liouvtotalfinal} \ea

We now notice that non-critical terms can be expressed in terms
of the Weyl anomaly coefficients for the
($\sigma$-model) graviton and dilaton backgrounds as:
\ba
&& 3\eta e^\Phi \frac{a'}{a} = 4\eta
e^{-\Phi}\left(\frac{1}{8}g^{\mu\nu}{\tilde \beta}_{\mu\nu}^{\rm
Grav} - \tilde \beta^\Phi e^{2\Phi}\right)\label{weylanom} \nonumber \\
&& 4\eta e^\Phi \Phi ' = 4\eta e^\Phi \tilde \beta^\Phi
\ea
where we used
the Einstein frame metric to contract indices, with $\tilde
\beta^{\rm Grav}$ denoting the graviton Weyl anomaly coefficient. In
(\ref{weylanom}) we have taken into account that in the class of
models of \cite{dglmn,dgmpp}, we are concentrating in this work,
$\tilde \beta_{00~~}^{\rm Grav} = 0$.

Thus, we observe from (\ref{liouvtotalfinal}), that terms proportional
to the dilaton $\tilde \beta^\Phi$  cancel out, leaving only the graviton
contributions as the non-critical string corrections to the
Boltzmann equation, which finally becomes:
\ba \frac{d n}{dt} + 3\left(\frac{\dot a}{a}\right)n - \dot \Phi n
= \frac{1}{2}\eta \left(e^{-\Phi}g^{\mu\nu}{\tilde \beta}_{\mu\nu}^{\rm Grav}\right)n
+ \int \frac{d^3p}{E}C[f]
\label{liouvtotalfinal2} \ea
It is important to recall once more that, upon Liouville dressing,
which restores the conformal invariance of the model, the graviton
and dilaton Weyl anomaly coefficients satisfy (\ref{liouveq}), which
upon the identification of the Liouville mode with (a function of
the cosmic) time yield (\ref{eqall}).

 We next remark that in the case with non-critical
 string backgrounds, consistency requirements between the two equations
determine the back reaction effects of matter onto space time. For
instance, in the absence of matter, in the specific (but generic
enough) non-critical string model of \cite{dgmpp}, we have that the
present-era dilaton and scale factor of the Universe are such that
${\dot Q} \simeq 0, ~\dot \Phi + \frac{\dot a}{a} = 0$. In this
particular case the off-shell terms at late eras are practically
zero, and the system has reached equilibrium. The corresponding
Boltzmann equation for the dilaton species densities is then
satisfied in agreement with the corresponding continuity equation
(\ref{contin}) in absence of matter.

However, in the presence of matter, we know from the analysis of
\cite{dglmn} that $\dot \Phi + \frac{\dot a}{a} \ne 0$, and the
right-hand-side of the Boltzmann equation (\ref{liouvtotal}), with
$n$ now representing matter species, yields terms proportional to
$n$. Consistency with the continuity equation (\ref{contin}), then,
implies that the off-shell (non-critical string background) terms in
the latter should be proportional to the total number density of
matter species (back reaction effects).  This and other similar
consistency checks should be made when considering solving such
non-equilibrium cosmologies.

\section{Phenomenology of Particle Physics Models and Modified Boltzmann equation}

 In this section we consider solutions of the
modified Boltzmann equation (\ref{liouvtotalfinal}), or equivalently
(\ref{liouvtotalfinal2}), for a particle species density $n$ in the
physically interesting case of supersymmetric dark matter species,
such as neutralinos, viewed as the lightest supersymmetric particles
(LSP). Such Cold Dark Matter candidates lead to a rich phenomenology
of supersymmetric particle physics models. In the context of
conventional Cosmology~\cite{susyconstr}, some of these models can
be constrained significantly by the recently available astrophysical
data on Cosmic Microwave Background temperature
fluctuations~\cite{wmap}. The calculation of relic abundances will
be done in some detail, in order for the reader to appreciate better
the r\^ole of the non conventional terms in
(\ref{liouvtotalfinal2}). .

It is convenient to write the Boltzmann equation for the density of
species $n$ in a compact form that represents collectively the
dilaton-dissipative-source and non-critical-string contributions as
external-source $\Gamma (t) n$ terms: \ba \frac{d n}{dt} +
3\frac{\dot a}{a}n = \Gamma (t)n + \int \frac{d^3p}{E}C[f] ~, \qquad
\Gamma (t) \equiv  \dot \Phi  + \frac{1}{2}\eta
e^{-\Phi}g^{\mu\nu}{\tilde \beta}_{\mu\nu}^{\rm Grav}
\label{boltzfinal}
 \ea
where we work in the physical scheme (\ref{liouv-time}) from now on,
for which $\eta = -1$. Depending on the sign of $\Gamma (t)$ one has
different effects on the relic abundance of the species $X$ with
density $n$, which we now proceed to analyze.
To find an explicit expression for $\Gamma (t)$ in our case we should
substitute the solution of (\ref{liouveq}), more specifically (\ref{eqall}),
analyzed in \cite{dglmn}. Regarding the form of (\ref{boltzfinal}) it is
nice to see that the extra terms can be cast in a simple-looking form of a source term
$\Gamma (t) n$ including both the dilaton dissipation and the non-critical-string terms.
Of course  $\Gamma (t)$ is complicated and requires the full solution outlined in \cite{dglmn}.

 In a more familiar form, the interaction term $C[f]$
of the above modified Boltzmann equation can be
expressed
in terms of the thermal average of the cross section $\sigma$
times the Moeller velocity $v$ of the annihilated particles~\cite{kolb}

\bear
\frac{dn}{dt}\;=\; - 3 \; \frac{\dot{a}}{a} \; n - \vev{v \sigma}\;( n^2 - n_{eq}^2) + \Gamma \;n
\label{bbb}
\eear
 Before the decoupling time $t_{f}$, $t < t_f$, equilibrium is
maintained and thus $n=n_{eq}$ for such an era. However, it is
crucial to observe that, as a result of the presence of the source
$\Gamma$ terms, $n_{eq}$ no longer scales with the inverse of the
cubic power  of the expansion radius $a$, which was the case in
conventional (on-shell) cosmological models .

To understand
this, let us assume that $n= n_{eq}^{(0)}$ at a very early epoch
$t_0$. Then the solution of the modified Boltzmann equations at all
times $t<t_f$ is given by
\bear
n_{eq} a^3 \;=\; n_{eq}^{(0)} \; a^3(t_0) \; \exp \;({ \; \int_{t_0}^{t} \Gamma dt })
\; . \label{equ}
\eear
 The time $t_0$ characterizes a very early time, which is
not unreasonable to assume that it signals the exit from the
inflationary period. Soon after the exit from inflation, all
particles are in thermal equilibrium, for all times $t<t_f$,  with
the source term modifying the usual Boltzmann distributions in the
way indicated in Eq. (\ref{equ}) above. It has been tacitly assumed
that the entropy is conserved despite the presence of the source
and the non-critical-string contributions. In our approach this is an
{\it approximation}, since we know that non-critical strings
lead to entropy production. However, as argued
in our previous works on the subject~\cite{emn}, the entropy increase
is most significant during the inflationary era, and hence it is
not inconsistent to assume that, for all practical purposes,
sufficient  for our phenomenological analysis
in this work,
there is no significant entropy production after the exit from inflation.
This is a necessary ingredient for our approach,
since without such an assumption no predictions can be
made, even in the conventional cosmological scenarios. Thus, the
picture we envisage is that at $t_0$ the Universe entered an equilibrium phase,
the entropy is conserved to
a good approximation,
and hence all particle species find themselves
in thermal
equilibrium, despite
the presence of the $\Gamma$ source, which \emph{slowly} pumps
in or sucks out energy, without, however, disturbing the particles' thermal
equilibrium.

From the above discussion it becomes evident that it is of
paramount importance to know the behaviour of the source term at all
times, in order to extract information for the relic abundances,
especially those concerning Dark Matter, and how these are
modified from those of the standard Cosmology. Before embarking on
such an  enterprize and study the phenomenological consequences of
particular models predicting the existence of Dark Matter,
especially Supersymmetry-based ones, we must first proceed in a
general way to set up the stage and discuss how
the relic density is affected by the presence of the
non conventional source
terms discussed above.

For the sake of brevity, we shall not deploy all the details of the
derivation of the relic density, but instead demonstrate the most
important features and results of our approach, paying particular
attention to exhibiting the differences from the conventional case.
Generalizing the standard techniques~\cite{kolb}, we assume that
above the freeze-out point the density is the equilibrium density as
provided by Eq. (\ref{equ}), while below this the interaction terms
starts becoming unimportant. It is customary to define $x \equiv
T/m_{\lsp}$ and restrict the discussion on a particular species
${\lsp}$ of mass $m_{\lsp}$, which eventually may play the role of
the dominant Dark Matter candidate. It also proves convenient to
trade  the number density $n$ for the quantity $Y \equiv n/s$, that
is the number per entropy density~\cite{kolb}. The equation for $Y$
is derived from (\ref{bbb}) and is given by
\bear
\frac{dY}{dx}= m_{\lsp} \vev{v \sigma}\; {( \;\frac{45}{\pi} G_N {\tilde g}_{eff}\;)}^{-1/2} \;( h+ \frac{x}{3} \frac{dh}{dx})\; ( Y^2-Y^2_{eq})
- \frac{\Gamma}{H x} \; ( 1+ \frac{x}{3 h} \frac{dh}{dx}) \; Y \quad .
\label{eq2}
\eear
where $G_N =1/M_{Planck}^2$ is the four-dimensional gravitational
constant, the quantity $H$ is the Hubble expansion rate, $h$ denote
the entropy degrees of freedom, and ${\vev{ v \sigma}}$ is the
thermal average of the relative velocity times the annihilation
cross section and ${\tilde{g}}_{eff}$ is simply defined by the
relation~\cite{kolb}
\bear\label{gtild}
\varrho+\Delta \varrho \;\equiv \; \frac{\pi^2}{30}\; T^4 \;{\tilde{g}}_{eff} \quad .
\eear
The reader should notice at this point that
$\Delta \varrho$ incorporates the effects of the additional
contributions due to the non-critical (off-shell) terms and the
dilaton dissipative source, which are not accounted for in the
$g_{eff}$ of conventional Cosmology~\cite{kolb}, hence the notation
${\tilde{g}}_{eff}$.  We next remark that 
$\rho$, as well as $\Delta \rho$, as 
functions of time are known, once one solves 
the cosmological equations.  However, only the degrees of freedom 
involved in $\rho$ are thermal,  the rest, like the comological-constant 
term if present
in a model, are included in 
$\Delta \rho$. Therefore, the relation 
between temperature and time is provided by
\bear
\rho= \frac{\pi^2}{30}\;T^4\;g_{eff}(T)
\label{timetemp}
\eear 
while $\rho + \Delta \rho$ are involved in the evolution through (c.f. (\ref{eqall}))
\bear
H^2=\frac{8 \pi G_N}{3}\; (\rho + \Delta \rho) \quad .
\label{ffff}
\eear
Thus, it is important for the reader 
to bear in mind that $\Delta \rho$ contributes to 
the dynamical expansion, through Eq. (\ref{ffff}), 
but not to the thermal evolution of the Universe. 
The quantity ${\tilde g}_{eff}$, defined in (\ref{gtild}),  
is therefore given by 
\bear
{\tilde g}_{eff}=g_{eff} + \frac{30}{\pi^2} T^{-4} \Delta \rho \quad .
\eear
The meaning of the above expression is that time has been replaced 
by temperature, through Eq. (\ref{timetemp}), after solving the 
dynamical equations. In terms of ${\tilde g}_{eff}$ the expansion 
rate $H$ is written as 
\bear\label{hubbleg}
H^2= \frac{4 \pi^3 G_N}{45} \; T^4 \;{\tilde g}_{eff} \quad .
\eear
This is used in the Boltzmann equation for $Y$ 
and the conversion from the time variable $t$ to temperature or, 
equivalently, the variable $x$.

  For $x$ above the freezing
point $x_f$, $Y \approx Y_{eq}$ and, upon omitting the contributions
of the derivative terms $dh/dx$, an approximation which is also
adopted in the standard cosmological treatments~\cite{kolb}, we
obtain for the solution of (\ref{eq2})
\bear Y_{eq} = Y_{eq}^{(0)} \; \exp \; ( \; - \int_{x}^{\infty}
\frac{\Gamma H^{-1}}{x} dx \;) \quad. \label{eq3} \eear
Here, $Y_{eq}^{(0)} $ corresponds to $n_{eq}^{(0)} $ and in the
non-relativistic limit is given by
\bear
Y_{eq}^{(0)} = \frac{45}{2 \pi^2} \frac{g_s}{h} \; {(2 \pi x)}^{-3/2} \exp {( -1/x )}
\label{bolit}
\eear
where $g_s$ counts the particle's spin degrees of freedom.

In the regime $x<x_f$, $Y >> Y_{eq}^{(0)}$ the equation (\ref{eq2})
can be written as
\bear
\frac{d}{dx} \frac{1}{Y}= - m_{\lsp} \vev{v \sigma}\; {( \;\frac{45}{\pi} G_N {\tilde g}_{eff}\;)}^{-\frac{1}{2}} h
+ \frac{\Gamma {H^{-1}}}{x Y} \; \quad
\label{eq22}
\eear
Applying (\ref{eq22}) at the freezing point $x_f$ and using
(\ref{eq3}) and (\ref{bolit}), leads, after a straightforward
calculation, to the determination of $x_f = T_f/m_{\lsp}$ through
\bear x_f^{-1}\;=\; ln \left[ 0.03824 \; g_s\; \frac{M_{Planck}~
m_{\lsp}}{\sqrt{g_{*}}} x_f^{1/2} {\vev{ v \sigma}}_f \right]\;+\;
\frac{1}{2} \;  ln \left( \frac{g_{*}}{ {\tilde{g}}_{*} }\right)
\;+\; \int_{x_f}^{x_{in}} \;  \frac{\Gamma H^{-1}}{x} \;dx \quad .
\label{xf} \label{frpo} \eear
As usual, all quantities are expressed in terms of the dimensionless
$x \equiv T/m_{\lsp}$ and $x_{in}$ corresponds to the time $t_0$
discussed previously, taken to represent the exit from the inflationary period
of the Universe.

 The first term on the right-hand-side of
(\ref{frpo}) is that of a conventional Cosmology for, say, an LSP
carrying $g_s$ spin degrees of freedom, playing the r\^ole of the
dominant Cold Dark Matter species in  a concrete and physically
promising example~\cite{susyconstr}, which we use in this work. The
quantity ${\vev{ v \sigma}}_f $ is the thermal average of $v \sigma$
at $x_f$ and $g_{*}$ is $g_{eff}$ of conventional Cosmology at the
freeze-out point. The same notation holds for ${\tilde{g}}_{*}$. In
our treatment above, we chose in (\ref{frpo}) to present $x_f$  in
such a way so as to separate the conventional contributions, which
reside in the first term, from the contributions of the dilaton and
the non-critical-string dynamics, which are contained within the
last two terms. The latter induce a shift in the freeze-out
temperature. The penultimate term on the right hand side of
(\ref{frpo}), due to its logarithmic nature, does not affect much
the freeze-out temperature. The last term, on the other hand, is
more important and, depending on its sign, may shift the freeze-out
point to earlier or later times. To quantify the amount of the shift
one must solve the equations (\ref{eqall}) of ref.~\cite{dglmn}.  We
shall do this in a forthcoming publication~\cite{spanos}, where we
shall also present a detailed analysis of the effects of the
non-critical string and dilaton-source terms on the constraints on
supersymmetric particle physics models, extending conventional
cosmology works~\cite{susyconstr}.

For our purposes here we note that, in order to calculate the
relic abundance, we must solve (\ref{eq22})
from $x_f$ to today's value $x_0$, corresponding to a temperature
$ T_0 \approx 2.7^0 K$.
Following the usual approximations we arrive at the result:
\bear
Y^{-1}(x_0)=Y^{-1}(x_f)+{(\frac{\pi}{45})}^{\frac{1}{2}}\; m_{\lsp} \; M_{Planck}\; {\tilde g}_{\star}^{-\frac{1}{2}}\; h(x_0) J -
\int_{x_0}^{x_f} \frac{\Gamma {H^{-1}}}{x Y} dx \quad .
\label{yoo}
\eear
In conventional Cosmology~\cite{kolb} ${\tilde g}_{\star}$ is
replaced by $g_{\star}$ and the last term in (\ref{yoo}) is absent. The
quantity $J$ is $J \equiv \int_{x_0}^{x_f} \vev{v \sigma} dx$. By
replacing $Y(x_f)$ by its equilibrium value (\ref{eq3}) the ratio of
the first term on the r.h.s. of (\ref{yoo}) to the second  is found
to be exactly the same as in the no-dilaton case. Therefore, by the
same token as in conventional Cosmology, the first term can be
safely omitted, as long as $x_f$ is  of order of $1/10$ or less.
Furthermore, the integral on the r.h.s. of (\ref{yoo}) can be
simplified if one uses the fact that  $\vev{v \sigma} n$ is small as
compared with the expansion rate $\dot{a}/a$ after decoupling . For
the purposes of the evaluation of this integral, therefore,  this
term can be omitted in (\ref{eq22}), as long as we  stay  within the
decoupling regime, and one obtains:
\bear
\frac{d}{dx} \frac{1}{Y}=    \frac{\Gamma {H^{-1}}}{x Y} \; \quad .
\label{eq222}
\eear
By integration this yields $Y(x)=Y(x_0) \; \exp( \; -\int_{x_0}^{x} \Gamma H^{-1} dx/x \; )$. Using this inside the integral in
(\ref{yoo}) we get
\bear
{(h(x_0) Y(x_0))}^{-1}= {\left( \;1+\int_{x_0}^{x_f} \frac{\Gamma {H^{-1}}}{\psi(x)} dx\; \right)}^{-1}\;
{(\frac{\pi}{45})}^{\frac{1}{2}}\; m_{\lsp} \; M_{Planck}\; {\tilde g}_{\star}^{-\frac{1}{2}}\; J \quad
\label{yoo2}
\eear
 where the function $\psi(x)$   is given by
$\psi(x) \equiv x \exp( \; -\int_{x_0}^{x} \Gamma H^{-1} dx/x \; )$.
With the exception of the prefactor on the r.h.s. of (\ref{yoo2}),
this is identical in form to the result derived in standard
treatments, if ${\tilde g}_{\star}$ is replaced by $g_{\star}$ and
the value    of $x_f$, implicitly involved in the integral $J$, is
replaced by its value found in ordinary treatments in which the
dilaton-dynamics and non-critical-string effects are absent.

The matter density of species $\lsp$ is then given by 
\bear
\rho_{\lsp}=f \;{\left( \frac{4\pi^3}{45} \right)}^{1/2}\; {\left( \frac{T_{\lsp}}{T_\gamma}  \right)}^3
\frac{T_\gamma^3}{M_{Planck}} \frac{\sqrt{\tilde g}_{*}}{J}
\eear
where the prefactor $f$ is: 
$$
f= 1 \;+\; \int_{x_{0}}^{x_f}  \frac{\Gamma H^{-1}}{\psi(x)}
$$
It is important to recall that the thermal degrees of freedom are 
counted by $g_{eff}$ (c.f. (\ref{timetemp})), and not ${\tilde g}_{eff}$,
the latter being merely
a convenient device connecting
the total energy, thermal and non-thermal, to the temperature $T$ (c.f. 
(\ref{gtild})).
Hence, 
\bear\label{tgs}
{\left( \frac{T_{\lsp}}{T_\gamma}  \right)}^3=\frac{g_{eff}(1 MeV)}{g_{eff}(T_{\lsp})}\;\frac{4}{11}=\frac{43}{11} 
\frac{1}{g_{*}}  \quad .
\eear
In deriving (\ref{tgs}) only the thermal content of the Universe is used,
while the dilaton and the non-critical terms do not participate. 
Therefore the ${\lsp}$'s matter  density is given by 
\bear
\rho_{\lsp}=f\;{\left( \frac{4\pi^3}{45} \right)}^{1/2}\; \frac{43}{11}
\frac{T_\gamma^3}{M_{Planck}} \frac{\sqrt{\tilde g}_{*}}{g_{*} \;J} \quad .
\eear
This formula tacitly assumes that the ${\lsp}$s decoupled before neutrinos. 
For the relic abundance, then, we derive the following
approximate result
\bear
\Omega_{\tilde{\chi}} h_0^2 \;=\; \left( \Omega_{\tilde{\chi}} h_0^2 \right)_{no-source}
\times {\left(  \frac{{\tilde g}_{*}}{g_{*}}   \right)}^{1/2} \;
\left( 1 \;+\; \int_{x_{0}}^{x_f}  \frac{\Gamma H^{-1}}{\psi(x)} dx \right) \quad . \label{relic}
\eear
The quantity referred to as {\em{ no-source}} is the well known no-source expression
\bear \left( \Omega_{\tilde{\chi}} h_0^2 \right)_{no-source}\;=\;\
\frac{1.066\; \times  10^9\; {\rm GeV}^{-1}}{ M_{Planck}
\;\sqrt{g_{*}} \;J} \label{final} \eear
where $J\equiv \int_{x_0}^{x_f} \vev{ v \sigma} dx$. However, as
already remarked, the end point $x_f$ in the integration is the
shifted freeze-out point as determined by Eq. (\ref{frpo}). The
merit of casting the relic density in such a form is that it clearly
exhibits the effect of the presence of the source.  Certainly, if an
accurate result is required, one can proceed without approximations
and handle the problem numerically as in the standard treatments. We
shall present such a more complete analysis in a forthcoming
publication~\cite{spanos}.

\section{Discussion: source effects on particle phenomenology}

 From the expression (\ref{relic}) above, it becomes
evident that the effects of the integral involving the source
$\Gamma$ on the relic density evolution may be quite important.
Indeed, if $\Gamma$ is kept negative at all times, this  results in
reduction of the relic density with time, contrary to what happens
in the case where $\Gamma $ is positive. In the former case,
predictions for supersymmetric models~\cite{susyconstr} can be
drastically altered, since the parameter space is enlarged, leaving
more room for supersymmetry, probably beyond the reach of LHC, even
in the case of constrained minimal supersymmetric standard models
with compact parameter spaces of the embedding minimal supergravity
theory . The opposite happens in the case of positive $\Gamma$,
where the parameter space is shrunk and predictions can be very
restrictive to almost excluding supersymmetry, especially if the
prefactor turns out to be a large number.

In order to get a rough picture of the
importance of such changes in the calculation of the relic density,
let us assume that $\dot{a}/a+\dot{\Phi}=0$, as would be the case in the
present era if matter were absent~\cite{dgmpp}.
 The non-critical terms would also
contribute little in this case and from (\ref{boltzfinal})
one would have $\Gamma
\approx \dot{\Phi}$. Therefore, $\Gamma H^{-1} \approx -1$ and the
function $\psi(x)$ would be $\psi(x) \approx x^2/x_0$. As a result,
the prefactor in (\ref{relic}) becomes $x_0/x_f$ which is an
enormously small number, of order $\sim 10^{-13}$ or so. Such a small number
would result practically  to no cosmological constraints on
supersymmetric models !  However this situation is not realized in
nature, since at the present era neither matter nor the non-critical
term contributions are negligible~\cite{dglmn}.
A decent approach is to solve the
cosmological equations (\ref{eqall})
and thus obtain the function $\Gamma (t)$ at all times,
$t$, from today to the remote past. This would allow for precise information
to be obtained on the value of the freezing point $x_f$ and of the
prefactor appearing in the relic density (\ref{relic}).
Thus, a complete
numerical treatment, along the lines presented in \cite{dglmn}, is needed
in
order to tackle important phenomenological questions regarding the
relic abundance of Dark Matter in this framework.

Nevertheless, the smallness of the
prefactor in the crude approximation we have employed above, gives us a
sneak preview of the drastic changes one may be faced with, as a
result of the non-critical string dynamics. Moreover, in such a framework
one could also tackle other important issues, such
as a possible resolution of the
gravitino overproduction
problem in effective supergravity inflationary models~\cite{grvino}.
In the non-critical
string framework discussed above, the gravitino, as a member of the
gravitational supermultiplet of the string, will feel the non-critical
and dilaton dissipation terms, while matter effects are unimportant
for this issue. Thus, its density may be substantially reduced, in a way
similar to that discussed above, as compared with conventional
supergravity scenarios~\cite{lmn2}.
\emph{affair \`a suivre...}

\section*{Acknowledgments}

The work of A.B.L. and N.E.M. is partially supported by funds made available
by the European Social Fund (75\%) and National (Greek)
Resources (25\%) - (EPEAEK II)PYTHAGORAS.
The work of D.V.N. is supported by D.O.E. grant DE-FG03-95-ER-40917.

\end{document}